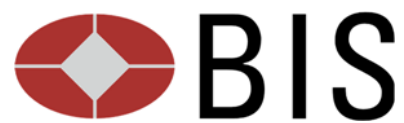

# BIS Working Papers
## No 1076

# The shape of business cycles: a cross-country analysis of Friedman's plucking theory

by Emanuel Kohlscheen, Richhild Moessner and Daniel M Rees

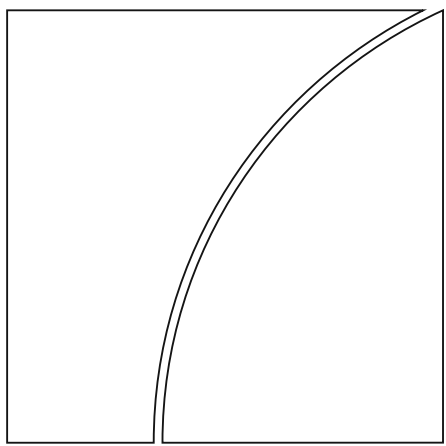

# The shape of business cycles: a cross-country analysis of Friedman's plucking theory

Emanuel Kohlscheen, Richhild Moessner and Daniel M. Rees [1]

## Abstract

We test the international applicability of Friedman's famous plucking theory of the business cycle in 12 advanced economies between 1970 and 2021. We find that in countries where labour markets are flexible (Australia, Canada, United Kingdom and United States), unemployment rates typically return to pre-recession levels, in line with Friedman's theory. Elsewhere, unemployment rates are less cyclical. Output recoveries differ less across countries, but more across episodes: on average, half of the decline in GDP during a recession persists. In terms of sectors, declines in manufacturing are typically fully reversed. In contrast, construction-driven recessions, which are often associated with bursting property price bubbles, tend to be persistent.

[1] Bank for International Settlements, Centralbahnplatz 2, 4002 Basel, Switzerland. We are grateful to Egon Zakrajšek for helpful comments and suggestions and Emese Kuruc for excellent research assistance. The views expressed in this paper are those of the authors and do not necessarily reflect those of the Bank for International Settlements.

## 1. Introduction

Are economic recoveries from recessions complete? If not, what characteristics influence the extent of recoveries? In this article, we present novel cross-country evidence on the shape of business cycles across advanced economies. We compare the patterns that are found with the predictions of Milton Friedman's "plucking theory" of the business cycle (see Friedman (1964, 1993)). This theory was recently revived by the influential paper of Dupraz *et al* (2019).[2]

Our analysis is based on twelve advanced economies (AEs), covering a total of more than 50 recessions since 1970.[3] Recessions in each country are timed using the same timing algorithm, following Bry and Boschan (1971) and Harding and Pagan (2003). Our central finding is that in countries with flexible labour markets post-recession declines in the unemployment rate almost fully reverse the effects of recessions. In other words, recoveries in these follow a fairly predictable pattern. Further, the unemployment rate declines in expansions bear no relation with rises in subsequent recessions. This pattern is broadly consistent with Friedman's plucking theory of the business cycle.

Yet, in the two thirds of countries with less flexible labour markets unemployment rates are far less cyclical. In contrast, output recoveries differ less across countries but more across episodes. Many recessions leave permanent scars. It is important to note here that, throughout, we focus on the *magnitude* of output and employment declines and subsequent recoveries, rather than looking at the *timing* or cause of turns in the business cycle.[4]

## 2. Dating business cycles across countries

Figure 1 shows the timeline of all economic expansions in twelve AEs since 1970. Expansions and recessions were timed by applying the Bry-Boschan-Harding-Pagan procedure to quarterly time series of seasonally adjusted real GDP.[5] For countries where commonly-accepted recession dates exist (as in the United States), the timeline aligns closely with those sources.

---

[2] Kim and Nelson (1999) state that in Friedman's plucking view, "recessions are like common cold: they come on suddenly and recovery follows a fairly predictable course, but the time that has passed since the last cold is of no use in predicting when the next will occur, or its severity."

[3] Australia, Canada, France, Germany, Italy, Japan, Norway, Spain, Sweden, Switzerland, the United Kingdom and the United States. Country selection was based solely on data availability (over long span).

[4] Whereas the overall state of an economy may not provide a hint of whether a new recession is likely, other indicators, such as the slope of the yield curve or downturns in residential construction activity have been shown to anticipate the *timing* of turning points. See e.g. Estrella and Hardouvelis (1991) and Kohlscheen *et al* (2020).

[5] All data were sourced from the OECD. For a complete explanation of the timing algorithm, see Harding and Pagan (2003). A minimum of two quarters per phase was imposed.

# Figure 1 – Timeline of economic expansions[1]

Peaks and throughs of business cycles as timed by Bry-Boschan-Harding-Pagan algorithm

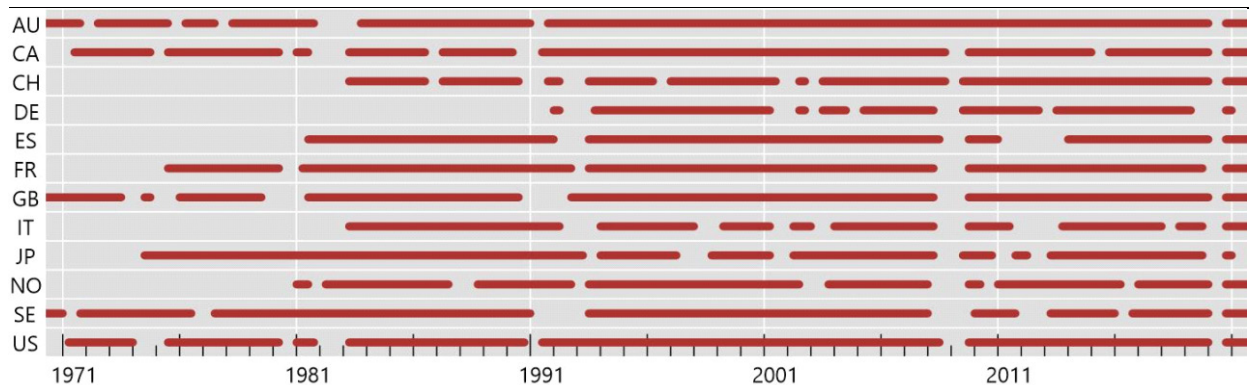

[1] The graph is showing expansion periods since Q1 1970, except for CA (Q3 1970), CH (Q3 1981), DE (Q3 1991), ES (Q4 1980), FR (Q3 1974), IT (Q4 1981), JP (Q4 1973) and NO (Q1 1980).

On average, we find that business cycles in advanced economies last 26 quarters (6½ years) – although this varies significantly across countries and over time. The identified dates also point to significant business cycle asymmetry: expansion phases last 23 quarters on average, while the average recession lasts for only 3½ quarters. Australia recorded the longest expansion in our sample, stretching for 114 quarters from Q2 1991 until Q4 2019, followed by Canada's 70 quarter expansion that started in Q1 1991 and ended in Q2 2008, with the global financial crisis.

## 3. The striking asymmetry in the business cycle

Figure 2 shows scatter plots of trough-to-peak changes in the unemployment rate versus preceding peak-to-trough changes and the corresponding fits. Each dot represents one full business cycle. Formally, we estimate the following regression,

$$\Delta u_{expansion,i} = c + \alpha \, \Delta u_{prev\ recession,i} + \epsilon_i \quad (1)$$

where $\Delta u_{expansion,i}$ is the change in the unemployment rate in country *i* in an expansion and $\Delta u_{prev\ recession,i}$ is the change in the unemployment rate in the prior recession, using OLS estimation with robust standard errors. We also estimate the following regression,

$$\Delta u_{recession,i} = c + \alpha \, \Delta u_{prev\ expansion,i} + \epsilon_i \quad (2)$$

where $\Delta u_{recession,i}$ is the change in the unemployment rate in a recession and $\Delta u_{prev\ expansion,i}$ is the change in the unemployment rate in the prior expansion. The regression results of equations (1) and (2) are shown in Table 1 and Figure 2 for all countries in the sample together, as well as separately for very flexible labour market countries, and for the remaining countries.

Larger increases in unemployment rates in recessions tend to go hand-in-hand with larger declines in subsequent expansions (columns (1-3) of Table 1, and Figure 2, left-hand panel). By contrast, unemployment rate increases in recessions bear no relationship to declines in the previous expansion (columns (4-6) of Table 1, and Figure 2, right-hand panel).

**Figure 2 – Labour market recoveries by degree of flexibility of labour markets**

In percentage points

Change in unemployment rate during expansion vs change in unemployment rate during previous recession

Change in unemployment rate in recession vs change in unemployment rate in previous expansion

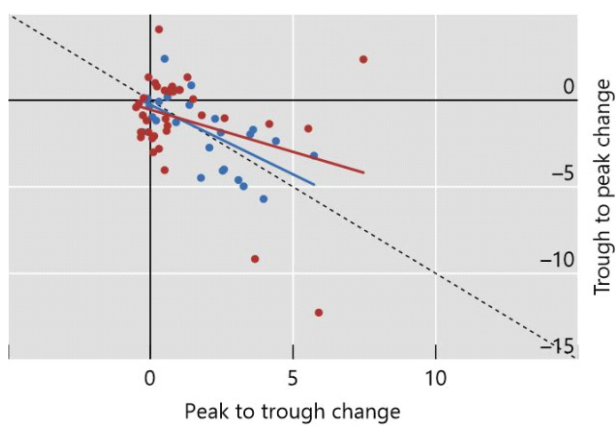
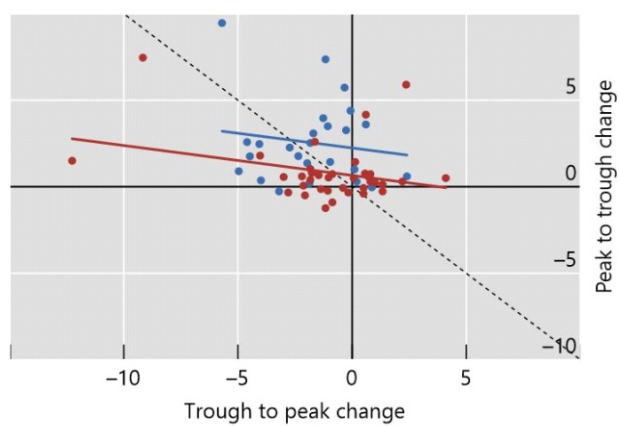

• CH, DE, ES, FR, IT, JP, NO, SE   • AU, CA, GB, US

Notes: Each dot represents one business cycle. For instance, peak to trough change stands for the change in the unemployment rate between the quarter of the peak of the GDP cycle, and the quarter of the trough of the GDP cycle in the respective country.

**Table 1: Unemployment cycles**

|  | (1) | (2) | (3) | (4) | (5) | (6) |
|---|---|---|---|---|---|---|
| Sample | all countries | very flexible labour market countries[a] | remaining countries[b] | all countries | very flexible labour market countries[a] | remaining countries[b] |
| Dependent variable | $\Delta u_{expansion}$ | $\Delta u_{expansion}$ | $\Delta u_{expansion}$ | $\Delta u_{recession}$ | $\Delta u_{recession}$ | $\Delta u_{recession}$ |
| Constant | -0.5050 (0.3574) | -0.2597 (0.4321) | -0.5445 (0.4003) | 1.2473*** (0.2782) | 2.2430*** (0.5437) | 0.6561** (0.2844) |
| $\Delta u_{prev\_recession}$ | -0.6045* (0.3144) | -0.8040*** (0.1917) | -0.4868 (0.4937) | - | - | - |
| $\Delta u_{prev\_expansion}$ | - | - | - | -0.2161 (0.1405) | -0.1700 (0.2900) | -0.1729 (0.1643) |
| No. of observations | 60 | 25 | 35 | 61 | 25 | 36 |
| Adjusted R² | 0.1671 | 0.3741 | 0.0760 | 0.0523 | -0.0205 | 0.0576 |

Notes: ***, ** and * represent significance at the 1%, 5% and 10% levels, respectively. OLS regressions with robust standard errors. [a] AU, CA, GB, US. [b] CH, DE, ES, FR, IT, JP, NO. SE.

Yet the evidence for complete unemployment recoveries is less clear-cut in a cross-country panel than it is from United States data alone. Dupraz *et al* (2019) had found that US unemployment rate increases during recessions were typically matched one-for-one by unemployment declines during expansions. In the pooled sample, the negative relationship is statistically significant only at the 10% significance level, A strong relationship practically holds for four countries in our sample: the United States, Canada, the United Kingdom and Australia (given the slope coefficient of −0.8). However, for the other eight countries, on average, only half of the increase in unemployment rates during recessions is reversed in expansions, and even this estimate is statistically insignificant at standard levels (slope −0.5).[6,7] The very weak relationship between falls in unemployment rates during expansions and increases in subsequent recessions holds for both sets of countries, however.

## 3.1. Interpretation

Several factors could explain the less complete unemployment recoveries in some countries. One possibility is that there is variation in the non-accelerating inflation rate of unemployment (NAIRU) that does not coincide with business cycle peaks and troughs. Indeed, several countries in our sample experienced large and persistent secular unemployment rate increases during the 1970s and 1980s, that in some cases persisted well into the 2010s.[8] Meanwhile, in some countries, unemployment rates display very little cyclical variation, even in deep recessions (e.g. Japan, Norway and Switzerland).

The above difference across country groups most likely reflects differences in labour market institutions. Indeed, it is notable that the countries in our sample that display the strongest evidence of complete unemployment recoveries are those with the most flexible labour markets (they have the four lowest average scores of the OECD Employment Protection Legislation Index).[9] Further, countercyclical policy responses that are too timid to fully offset economic downturns could also contribute to a less-than-complete reversal of unemployment rate increases, just as well-crafted policy settings could help to bring unemployment down quickly after recessions and limit the extent of overheating in expansions.

## 4. Output changes

In light of the above results, a natural question is whether cross-country variation in cyclical unemployment rate behaviour translates into different patterns of aggregate GDP. Identifying asymmetries in output is more complicated than for unemployment. Because GDP tends to increase over time, output losses during recessions are made up in all but the shortest expansions. And the total increase in GDP during an expansion clearly has more to do with its length than with the size of the previous recession.

---

[6] We cannot reject the null hypothesis that the slope coefficients are identical, however.

[7] This result continues to hold if we allow for the possibility that unemployment rates lag the business cycle by one or two quarters.

[8] See Blanchard and Summers (1986).

[9] OECD, 2020.

To abstract from these trend increases, we measure the cyclical component of GDP using the modification of the one-sided Hamilton filter proposed by Quast and Wolters (Hamilton (2018), Quast and Wolters (2022)). This filter measures the cyclical component of GDP as the residual from the regression $y_{t+h} = \frac{1}{9}\sum_{n=4}^{12} \hat{\theta}_{t,n}$ with $\hat{\theta}_{t,n} = y_t - \beta_{0,n} - \sum_{i=1}^{L} \beta_{i,n} y_{t-n-i+1}$ and where $y_t$ is the log of GDP.[10] We set L = 4. We use a one-sided version of the filter (i.e. for each quarter we use only information up to time *t* to estimate the regression coefficients). Our results are qualitatively similar if we use alternative cycle measures (e.g. a one-sided Hodrick-Prescott filter).

By construction, the filter measure will suggest that declines in output below trend during recessions tend to be reversed during expansions. To avoid overstating the extent to which lost output in recessions is reversed, we also examine the effect of recessions on trend output. We measure this trend as the difference between the five-year ahead forecast of GDP made at the business cycle peak and the two-year ahead forecast made three years after the peak. Intuitively, if recessions have purely transitory effects, medium-run output forecasts made before and after the recession should not systematically differ.

We estimate the following regressions,

$$\Delta y_{\text{expansion},i} = c + \alpha\, \Delta y_{\text{prev recession},i} + \epsilon_i \qquad (3)$$

$$\Delta y_{\text{recession},i} = c + \alpha\, \Delta y_{\text{prev expansion},i} + \epsilon_i \qquad (4)$$

$$\text{trend\_gr}_{\text{expansion},i} = c + \alpha\, \Delta y_{\text{prev recession},i} + \epsilon_i \qquad (5)$$

where the variables in equations (3) and (4) are the corresponding variables to those in equations (1) and (2), but for the cyclical component of output rather than for the unemployment rate. Moreover, trend_gr$_{\text{expansion}}$ is trend output growth in an expansion. The results of equations (3) to (5) are shown in Table 2 and Figure 3 for the full sample of Figure 1.

What we find is that output recoveries differ much less across countries than labour market recoveries. The relationship between the decline in output in downturns and its cyclical recovery in subsequent expansions mirrors that of the unemployment rate (coefficient –0.8, column (1) of Table 2 and Figure 3, left-hand panel). However, there is somewhat stronger evidence that larger output expansions coincide with deeper subsequent recessions (column (2) of Table 2 and Figure 3, centre panel). That said, the relationship is still considerably weaker than that between recession depth and subsequent output recoveries. In sum, output cycles across countries are also asymmetric, although less so than for those of unemployment.

---

[10] Intuitively, rather than measuring the cyclical component of GDP as the 8-quarter ahead forecast error of an AR(4) model, the Quast and Wolters filter calculates the cyclical component of GDP as the average of the forecast errors of an AR(4) model over an horizon of 4 to 12 quarters.

**Figure 3 – Output cycles**

In per cent

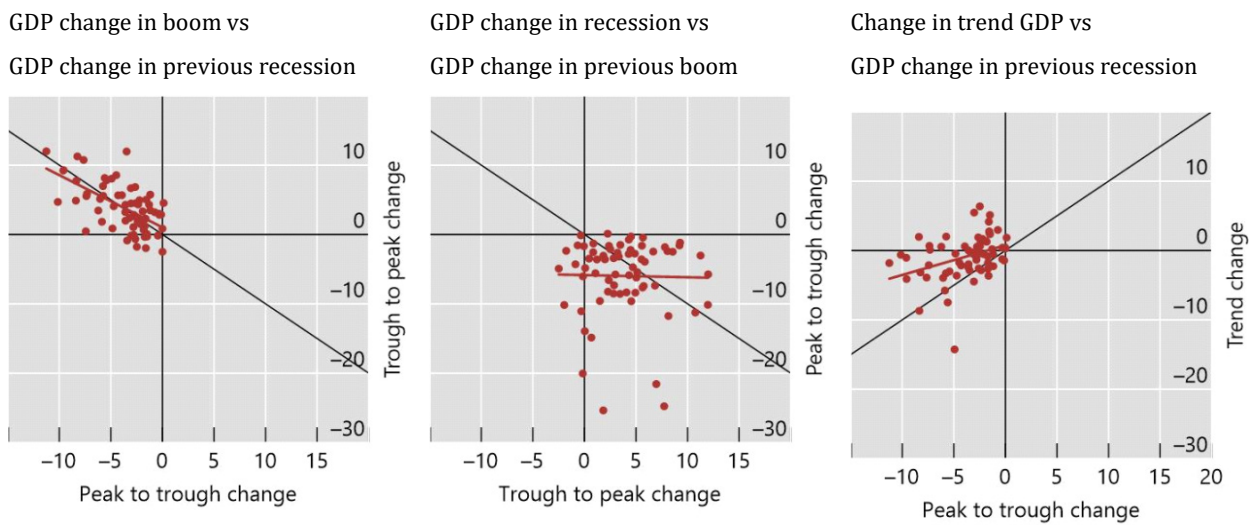

| GDP change in boom vs GDP change in previous recession | GDP change in recession vs GDP change in previous boom | Change in trend GDP vs GDP change in previous recession |

Notes: Each dot represents one business cycle (from Figure 1).

**Table 2: Output cycles**

|  | (1) | (2) | (3) |
|---|---|---|---|
| Dependent variable | $\Delta y_{expansion}$ | $\Delta y_{recession}$ | $trend\_gr_{expansion}$ |
| Constant | 1.0268** | -5.8540*** | 0.7471 |
|  | (0.5060) | (0.9700) | (0.4893) |
| $\Delta y_{prev\_recession}$ | -0.7533*** | - | 0.4272*** |
|  | (0.1175) |  | (0.1222) |
| $\Delta y_{prev\_expansion}$ | - | -0.0031 | - |
|  |  | (0.1960) |  |
| No. of observations | 68 | 69 | 67 |
| Adjusted $R^2$ | 0.3579 | -0.0148 | 0.1187 |

Notes: ***, ** and * represent significance at the 1%, 5% and 10% levels, respectively. OLS regressions with robust standard errors. Sample as shown in Figure 1: AU, CA, GB, US, CH, DE, ES, FR, IT, JP, NO. SE.

Recessions often have long-run output costs, however. This may occur through the reduction of potential output, due to lower capital investment or lower labour force participation - as pointed out by the hysteresis literature. For instance, Fernald et al (2017) and Yagan (2019), find that large unemployment shocks still had sizable negative effects on working age employment rates even many years after the GFC shock struck. This is confirmed also by Luo et al (2021)'s analysis of U.S. recessions, who conclude that employment rates have followed a L-shape rather than a U-shape after recent recessions.

What we find, is that internationally and on average, a bit under half of the decline in output during a recession shows up as a decrease in its medium-run trend (column (3) of Table 2 and regression line in right panel of Figure 3). That said, this average response masks a great deal of variation. In some recessions, almost all of the decline in output passes into a lower trend. But in around a quarter of recessions, the level of trend output is actually higher five years out. These contrasting effects are likely to reflect the fact that recessions can have many different causes, with those involving financial crises having particularly large and persistent consequences (Cerra and Saxena (2008)).

Our findings appear to be independent of monetary policy regimes and the length of cycle phases. Specifically, we find similar results when the sample is divided in pre- and post-1990 samples, i.e. before and after the widespread adoption of inflation targeting regimes. This also suggests that the average level of inflation, which was much higher pre-1990, does not affect them. Similarly, we find the same results when we divide the sample into one containing only recessions that were shorter and in one containing only recessions that were longer than the duration of the median recession. That said, the historical stability of these relationships does not guarantee that they will continue to hold for all possible policy regimes.[11]

## 5. Plucking pattern in manufacturing and trade, but not in construction

Industry-level data can shed further light on the sources of the business cycle asymmetry. Manufacturing cycles are highly asymmetric. Declines in recessions are typically fully made up (the slope coefficient in Figure 4, left panel is very close to –1), while the size of booms again tells us little about the depth of subsequent busts (Figure 4, centre panel). Cycles in services industries, such as retail trade, tend to have similar characteristics to those of aggregate output. These industries contribute to the plucking-type dynamics that we observe in aggregate economic activity.

The construction sector differs markedly from the other sectors. Increases in construction activity during expansions are almost completely unrelated to the depth of preceding recessions (the slope coefficient in the left-hand panel is statistically indistinguishable from zero). Moreover, the size of this industry's growth in expansions is strongly linked to its contractions in subsequent recessions. Instead of a plucking pattern, construction features boom-bust dynamics.

Several factors could explain the unusual cyclical behaviour of the construction industry. The industry is typically highly reliant on external financing, which could be plentiful at business cycle peaks, stimulating construction activity, and then dry up in recessions, leading to a bust. Construction activity is also closely tied to house price dynamics, which are prone to periods of speculative excess and subsequent periods of stagnation. Poor house price prospects after a boom reduce builders' effective rate of return for new construction and tends to depress construction activity for a protracted period.

---

[11]   Graphs with the results for these subsamples can be obtained upon request.

**Figure 4 – Regression coefficients by sector**

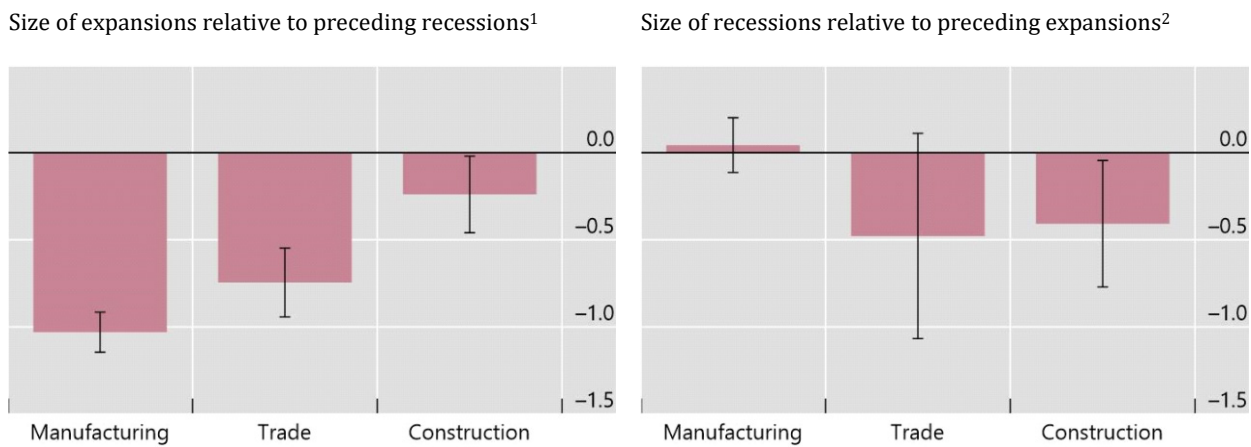

Size of expansions relative to preceding recessions[1]   Size of recessions relative to preceding expansions[2]

[1] Estimate of $\beta_1$ from a regression of $e_i = \beta_0 + \beta_1 r_i + \epsilon_i$ where $e_i$ is the value of cyclical gross value added in industry $i$ at the peak of an expansion and $r_i$ is the value of cyclical gross value added in industry $i$ at the trough of the preceding recession. Cyclical components of gross value added calculated using a Hamilton Filter with a lag length of 4 quarters and a horizon of 8 quarters. The countries and recession dates used in the regression are as listed in Graph 1. Gross value added data is not available for all industries and countries, particularly early in the data sample.   [2] Estimate of $\beta_1$ from a regression of $r_i = \beta_0 + \beta_1 e_i + \epsilon_i$.

## 6. Conclusion

While large recessions tend to precede strong expansions, the opposite is not typically the case. This observation, based on a sample of twelve AEs and more than fifty recessions, is consistent with Friedman's plucking theory of the business cycle. Yet evidence for the plucking theory is considerably stronger for the United States than it is for a broader set of AEs, and is more evident in the unemployment rate than in GDP. Recessions often lead to a permanent decline in the level of GDP. The source of the decline in output also matters. Most notably, activity in the construction industry looks to be strongly at odds with the predictions of the plucking theory.

Our results suggest that policymakers should aim to restore full employment quickly after a recession, while monitoring if their actions do not create excessive activity in the construction sector – where hysteresis is large and excesses are costly.

The economic recovery from the pandemic recession has been faster than initially anticipated in most AEs. A factor that has been fundamental in this respect is that the recession was triggered by a truly exogenous shock, rather than financial excesses or aggressive policy tightening to contain inflation. Moreover, the shock did not originate in the construction sector, where recoveries are much slower and partial, especially if associated with the bursting of a property price bubble. Instead, manufacturing and international trade in goods proved to be surprisingly resilient, given the extent of the adverse shock. Policy support was quick and unprecedented in size, decisively arresting the fall in employment and demand.

This is not to say that a full recovery is a given. Sectoral reallocation remains a challenge, particularly when skills required in expanding sectors do not match those that become available from the shrinking sectors. Continuous re-skilling and up-skilling of the labour force is also central for maintaining a flexible economy, which can adjust more rapidly to shocks of different kinds.

**Table A1 - Recession timeline**

| | | | duration | | unemployment rate | | | output index | | |
|---|---|---|---|---|---|---|---|---|---|---|
| cty | peak | trough | recession | expansion | cycle peak | cycle trough | subsequent peak | cycle peak | cycle trough | subsequent peak |
| AU | 1971Q3 | 1972Q1 | 2 | 22 | 1.9 | 2.4 | 4.8 | 26.7 | 26.3 | 29.5 |
| AU | 1975Q2 | 1975Q4 | 2 | 13 | 4.8 | 5.4 | 5.6 | 29.5 | 28.8 | 30.9 |
| AU | 1977Q2 | 1977Q4 | 2 | 6 | 5.6 | 5.9 | 5.8 | 30.9 | 30.7 | 35.5 |
| AU | 1981Q3 | 1983Q2 | 7 | 15 | 5.8 | 10.2 | 7.9 | 35.5 | 34.2 | 46.3 |
| AU | 1990Q4 | 1991Q2 | 2 | 30 | 7.9 | 9.6 | 5.2 | 46.3 | 45.7 | 111.2 |
| AU | 2019Q4 | 2020Q2 | 2 | 114 | 5.2 | 6.9 | 4.9 | 111.2 | 103.1 | 112.9 |
| CA | 1974Q3 | 1975Q1 | 2 | 14 | 5.3 | 6.7 | 7.6 | 35.8 | 35.4 | 43.4 |
| CA | 1980Q1 | 1980Q3 | 2 | 20 | 7.6 | 7.5 | 7.2 | 43.4 | 43.3 | 45.4 |
| CA | 1981Q2 | 1982Q4 | 6 | 3 | 7.2 | 12.9 | 9.7 | 45.4 | 43 | 50.8 |
| CA | 1986Q2 | 1986Q4 | 2 | 14 | 9.7 | 9.5 | 7.6 | 50.8 | 50.5 | 56.9 |
| CA | 1990Q1 | 1991Q1 | 4 | 13 | 7.6 | 10.2 | 6.1 | 56.9 | 55 | 90.6 |
| CA | 2008Q3 | 2009Q2 | 7 | 70 | 6.1 | 8.6 | 6.7 | 90.6 | 86.5 | 100.6 |
| CA | 2014Q4 | 2015Q2 | 2 | 22 | 6.7 | 6.9 | 5.7 | 100.6 | 99.7 | 109.2 |
| CA | 2019Q4 | 2020Q2 | 2 | 18 | 5.7 | 13.1 | 7.8 | 109.2 | 94.9 | 107 |
| CH | 1986Q2 | 1986Q4 | 2 | 14 | 1.5 | 1.4 | 1.1 | 59.3 | 59.2 | 67.5 |
| CH | 1990Q2 | 1991Q2 | 4 | 14 | 1.1 | 1.6 | 2.5 | 67.5 | 66.4 | 67.1 |
| CH | 1992Q1 | 1993Q1 | 4 | 3 | 2.5 | 4.4 | 4.9 | 67.1 | 65.6 | 68.1 |
| CH | 1996Q1 | 1996Q3 | 2 | 12 | 4.9 | 5.4 | 2.2 | 68.1 | 67.7 | 77.2 |
| CH | 2001Q2 | 2002Q1 | 3 | 19 | 2.2 | 2.8 | 3.3 | 77.2 | 76.9 | 77.2 |
| CH | 2002Q3 | 2003Q1 | 2 | 2 | 3.3 | 4.0 | 3.2 | 77.2 | 76.3 | 91.4 |
| CH | 2008Q3 | 2009Q1 | 2 | 22 | 3.2 | 3.7 | 4.1 | 91.4 | 87.4 | 108.7 |
| DE | 1992Q1 | 1993Q2 | 5 | 2 | 7.5 | 8.7 | 9.3 | 82 | 80.3 | 93.8 |
| DE | 2001Q1 | 2002Q1 | 4 | 31 | 9.3 | 9.6 | 9.8 | 93.8 | 93 | 93.9 |
| DE | 2002Q3 | 2003Q1 | 2 | 2 | 9.8 | 10.5 | 10.5 | 93.9 | 92.4 | 93.7 |
| DE | 2004Q2 | 2004Q4 | 2 | 5 | 10.5 | 10.8 | 8.0 | 93.7 | 93.5 | 102.8 |
| DE | 2008Q1 | 2009Q1 | 4 | 13 | 8.0 | 7.9 | 6.8 | 102.8 | 95.6 | 104.9 |
| DE | 2012Q3 | 2013Q1 | 2 | 14 | 6.8 | 6.9 | 5.0 | 104.9 | 104 | 117.2 |
| DE | 2019Q1 | 2020Q2 | 5 | 24 | 5.0 | 6.1 | 6.2 | 117.2 | 103.5 | 113.7 |
| ES | 2010Q4 | 2013Q3 | 11 | 6 | 20.2 | 26.1 | 13.8 | 100.2 | 94.9 | 111.4 |
| ES | 2019Q4 | 2020Q2 | 2 | 25 | 13.8 | 15.4 | 15.4 | 111.4 | 86.7 | 103.9 |
| FR | 1980Q1 | 1980Q4 | 3 | 20 | 5.5 | 5.8 | 10.8 | 53.5 | 53.1 | 69.9 |
| FR | 1992Q3 | 1993Q1 | 2 | 47 | 10.8 | 11.3 | 7.3 | 69.9 | 69.3 | 96.7 |
| FR | 2008Q1 | 2009Q2 | 5 | 60 | 7.3 | 9.1 | 8.2 | 96.7 | 93 | 107.7 |
| FR | 2019Q3 | 2020Q2 | 3 | 41 | 8.2 | 7.3 | 7.5 | 107.7 | 87.7 | 103.3 |
| GB | 1973Q2 | 1974Q1 | 3 | 32 | 3.7 | 3.6 | 3.7 | 42 | 40.3 | 41.1 |
| GB | 1974Q3 | 1975Q3 | 4 | 2 | 3.7 | 4.7 | 5.3 | 41.1 | 39.7 | 46.6 |
| GB | 1979Q2 | 1981Q1 | 7 | 15 | 5.3 | 8.9 | 6.7 | 46.6 | 44.1 | 60.2 |
| GB | 1990Q2 | 1992Q2 | 8 | 37 | 6.7 | 9.8 | 5.2 | 60.2 | 59 | 94 |
| GB | 2008Q1 | 2009Q2 | 5 | 63 | 5.2 | 7.8 | 3.8 | 94 | 88.5 | 106.6 |
| GB | 2019Q4 | 2020Q2 | 2 | 42 | 3.8 | 4.1 | 4.7 | 106.6 | 83.4 | 102 |
| IT | 1992Q1 | 1993Q3 | 6 | 37 | 8.7 | 10.0 | 11.3 | 87.2 | 85.9 | 95.3 |
| IT | 1997Q4 | 1998Q4 | 4 | 17 | 11.3 | 11.4 | 9.4 | 95.3 | 94.9 | 102.6 |
| IT | 2001Q1 | 2001Q4 | 3 | 9 | 9.4 | 8.9 | 8.5 | 102.6 | 101.9 | 103 |
| IT | 2002Q4 | 2003Q3 | 3 | 4 | 8.5 | 8.4 | 6.6 | 103 | 102.4 | 109.4 |
| IT | 2008Q1 | 2009Q2 | 5 | 18 | 6.6 | 7.6 | 8.2 | 109.4 | 101.3 | 104.7 |
| IT | 2011Q2 | 2013Q2 | 8 | 8 | 8.2 | 12.4 | 11.0 | 104.7 | 99.1 | 103.9 |
| IT | 2017Q4 | 2018Q2 | 2 | 18 | 11.0 | 10.8 | 9.7 | 103.9 | 103.8 | 104.4 |
| IT | 2019Q3 | 2020Q2 | 3 | 5 | 9.7 | 8.5 | 10.1 | 104.4 | 85.3 | 100.1 |
| JP | 1997Q1 | 1998Q2 | 5 | 14 | 3.3 | 4.1 | 4.8 | 89.1 | 87.2 | 91.2 |
| JP | 2001Q1 | 2001Q4 | 3 | 11 | 4.8 | 5.4 | 3.9 | 91.2 | 89.2 | 98.3 |
| JP | 2008Q1 | 2009Q1 | 4 | 25 | 3.9 | 4.6 | 5.1 | 98.3 | 89.6 | 96.3 |
| JP | 2010Q3 | 2011Q2 | 3 | 6 | 5.1 | 4.7 | 4.5 | 96.3 | 93.7 | 97.1 |
| JP | 2012Q1 | 2012Q4 | 3 | 3 | 4.5 | 4.2 | 2.3 | 97.1 | 95.8 | 103.6 |
| JP | 2019Q3 | 2020Q2 | 3 | 27 | 2.3 | 2.7 | 3.0 | 103.6 | 93 | 100 |
| NO | 1981Q2 | 1981Q4 | 2 | 3 | 2.1 | 2.0 | 1.9 | 47.4 | 46.8 | 59.1 |
| NO | 1987Q2 | 1988Q2 | 4 | 22 | 1.9 | 3.0 | 5.9 | 59.1 | 57.5 | 64.3 |
| NO | 1992Q3 | 1993Q1 | 2 | 17 | 5.9 | 6.0 | 3.9 | 64.3 | 63.5 | 89.4 |
| NO | 2002Q2 | 2003Q2 | 4 | 37 | 3.9 | 4.5 | 2.3 | 89.4 | 88.7 | 102.1 |
| NO | 2007Q4 | 2009Q2 | 6 | 18 | 2.3 | 3.3 | 4.0 | 102.1 | 99.3 | 101.4 |
| NO | 2010Q1 | 2010Q3 | 2 | 3 | 4.0 | 3.7 | 4.9 | 101.4 | 97.8 | 110.1 |
| NO | 2016Q1 | 2016Q3 | 2 | 22 | 4.9 | 4.8 | 4.0 | 110.1 | 108.4 | 116.8 |
| NO | 2019Q4 | 2020Q2 | 2 | 13 | 4.0 | 4.6 | 5.1 | 116.8 | 109.8 | 116.1 |
| SE | 1990Q4 | 1993Q1 | 9 | 15 | 2.2 | 7.7 | 6.1 | 59.5 | 55.9 | 91.5 |
| SE | 2007Q4 | 2009Q3 | 7 | 59 | 6.1 | 8.7 | 7.6 | 91.5 | 85.9 | 94.8 |
| SE | 2011Q3 | 2012Q4 | 5 | 8 | 7.6 | 8.2 | 7.1 | 94.8 | 93.1 | 102.9 |
| SE | 2015Q4 | 2016Q2 | 2 | 12 | 7.1 | 6.9 | 7.0 | 102.9 | 102.8 | 110.9 |
| SE | 2019Q4 | 2020Q2 | 2 | 14 | 7.0 | 8.5 | 9.2 | 110.9 | 101.4 | 110.8 |
| US | 1969Q3 | 1970Q4 | 5 | 46 | 3.6 | 5.8 | 4.8 | 28.5 | 28.3 | 32.9 |
| US | 1973Q4 | 1975Q1 | 5 | 12 | 4.8 | 8.3 | 6.3 | 32.9 | 31.8 | 39.2 |
| US | 1980Q1 | 1980Q3 | 2 | 20 | 6.3 | 7.7 | 7.4 | 39.2 | 38.4 | 40 |
| US | 1981Q3 | 1982Q4 | 5 | 4 | 7.4 | 10.7 | 5.7 | 40 | 39 | 53.9 |
| US | 1990Q3 | 1991Q1 | 2 | 31 | 5.7 | 6.6 | 5.3 | 53.9 | 53.2 | 90.4 |
| US | 2008Q2 | 2009Q2 | 4 | 69 | 5.3 | 9.3 | 3.6 | 90.4 | 86.8 | 110.5 |
| US | 2019Q4 | 2020Q2 | 2 | 42 | 3.6 | 13.1 | 5.9 | 110.5 | 99.3 | 111.3 |

# Previous volumes in this series

All volumes are available on our website www.bis.org.